# The funding effect on citation and social attention: the UN Sustainable Development Goals (SDGs) as a case study

Pablo Dorta-González [1,*] and María Isabel Dorta-González [2]

[1] University of Las Palmas de Gran Canaria, TiDES Research Institute, Campus de Tafira, 35017 Las Palmas de Gran Canaria, Spain. E-mail: pablo.dorta@ulpgc.es ORCID: http://orcid.org/0000-0003-0494-2903

[2] University of La Laguna, Departamento de Ingeniería Informática y de Sistemas, Avenida Astrofísico Francisco Sánchez s/n, 38271 La Laguna, Spain. E-mail: isadorta@ull.es

* Corresponding author

## Abstract

*Purpose:* Academic citation and social attention measure different dimensions in the impact of research results. We quantify the contribution of funding to both indicators considering the differences attributable to the research field and access type.

*Design/methodology/approach:* Citation and social attention accumulated until the year 2021 of more than 367 thousand research articles published in the year 2018, are studied. We consider funding acknowledgements in the research articles. The data source is Dimensions and the units of study are research articles in the UN Sustainable Development Goals.

*Findings:* Most cited goals by researchers do not coincide with those that arouse greater social attention. A small proportion of articles accumulates a large part of the citations and most of the social attention. Both citation and social attention grow with funding. Thus, funded research has a greater probability of being cited in academic articles and mentioned in social media. Funded research receives on average two to three times more citations and 2.5 to 4.5 times more social attention than unfunded research. Moreover, the open access modalities gold and hybrid have the greatest advantages in citation and social attention due to funding.

*Originality:* The joint evaluation of the effect of both funding and open access on social attention.

*Research limitations:* Specific topics were studied in a specific period. Studying other topics and/or different time periods might result in different findings.

*Practical implications:* When funding to publish in open or hybrid access journals is not available, it is advisable to self-archiving the pre-print or post-print version in a freely accessible repository.

*Social implications:* Although cautiously, it is also advisable to consider the social impact of the research to complement the scientific impact in the evaluation of the research.

**Keywords:** funding acknowledgements, funded research, citation, social attention, altmetric, open access, UN Sustainable Development Goals, SDGs

**Introduction**

The acknowledgments section in a research publication expresses the gratitude of the authors to those entities that contributed, inspired or financed the research (Costas and van Leeuwen, 2012). The present study focuses on a specific type of acknowledgements, the funding acknowledgements (FAs). Funding represents an important input in the research process. For this reason, it is not surprising that the topic of FAs has interested researchers since the 1970s (Álvarez-Bornstein and Montesi, 2020; Desrochers *et al.*, 2016, 2017; Liu, 2020; Liu *et al.*, 2020).

The paper by Costas and van Leeuwen (2012) is one of the first bibliometric studies about FAs using information on a large scale. These authors analyzed the information related to FAs in scientific publications of the year 2009 in the Web of Science database. They studied the association between the impact of the research and factors such as the scientific discipline, the country of authors, the documentary typology, and the type of collaboration. They found that 43% of the publications declare FAs, with significant variability between countries. They also observed that publications with AFs are more cited.

However, citations are only one side of a multidimensional concept such as research impact, and different alternatives have been considered in the literature to complement the impact of research funding. Zhao *et al.* (2018) analyzed the association between usage count and FAs in more than 300 thousand articles published in 2013 in six subject categories. They concluded that a positive correlation between funding and usage metrics exists, but with differences among disciplines. On this context, Álvarez-Bornstein and Montesi's (2020) systematic review of the literature concluded that advances of research into FAs might be achieved in the topic of societal impact of research.

Governments increasingly push researchers toward activities with social impact, including economics, cultural and health benefits (Thelwall, 2021). Thus, since the term 'altmetrics' was introduced in 2010 (Priem *et al.*, 2010), theoretical and practical scientific investigations were conducted in this discipline (Sugimoto *et al.*, 2017).

Most altmetric data improve citations in terms of the accumulation speed after publication (Fang and Costas, 2020; Williams, 2017). Thus, scholars analyzed the correlation between altmetrics and citations (Banshal *et al.*, 2021; Thelwall and Nevill, 2018).

However, except for Mendeley readership which is moderately correlated with citations (Zahedi and Haustein, 2018), there is a negligible or weak correlation between citations and most altmetrics (Banshal *et al.*, 2021; Bornmann, 2015; Costas *et al.*, 2015). This means that altmetrics might capture other forms of impact rather than citation impact (Wouters *et al.*, 2019). Scholars studied the differences between citations and altmetrics for finnish articles (Didegah *et al.*, 2018). The distribution of impact metrics is skewed for both citations and altmetrics (Banshal *et al.*, 2022), and the disciplinary aspect plays a relevant role for both the citations and altmetric scores (Banshal *et al.*, 2019; Htoo and Na, 2017).

There are four main Open Access (OA) modalities. OA Gold refers to papers in freely accessible OA journals. OA Green refers to publishing in a pay-per-view journal, in addition to self-archiving the pre-print or post-print paper in a freely accessible repository. OA Hybrid is an intermediate modality where authors pay publishers to make papers freely accessible within pay-per-view journals. And OA Bonze refers to papers in pay-per-view journals made available freely accessible by the publisher for some period without guarantees of permanence.

Some of the OA impact advantage is likely due to more access allows more people to read articles they otherwise would not. However, causation is difficult to establish and there are many possible biases (Dorta-González *et al.*, 2017; Dorta-González and Santana-Jiménez, 2018; Dorta-González and Dorta-González, 2022). Several factors can affect the observed differences in impact and funder mandates can be one of them. Funders are likely to have OA requirement, and well-funded studies are more likely to receive more citations than poorly funded studies (Aagaard *et al.*, 2020).

We analyze in this paper the effect of both funding and open access on social attention. There are other studies that evaluated the effect of funding on social attention (Doğramaci and Rossi-Fedele, 2022) or the effect of open access on social attention (Yu *et al.*, 2022). However, the novelty in our paper is the joint evaluation of both funding and open access on social attention.

Thus, based on citation and social attention data from the Dimensions database, we provide transversal estimations of the funding advantage in all publication modalities: OA Gold, OA Hybrid, OA Bronze, OA Green, and Closed. For it, we consider the accumulated citation and social attention until the year 2021 of a total of 367,704 research articles published in the year 2018. The data source is Dimensions and the research articles analyzed correspond to all those classified in this database in the 17 UN Sustainable Development Goals (SDGs).

**Literature review**

A central question in social studies of science is how to improve the dissemination of scientific results to increase societal impact. This aspect has been studied by many scholars, such as Li *et al.* (2013) who focused on the role of the authorship network on research impact. Other studies analyzed the relationship between journal impact factor and article citation (Campanario *et al.*, 2011; Larivière and Gingras, 2010; Thelwall & Wilson, 2014).

Scholars also analyzed the effect of funding on citation impact (MacLean *et al.*, 1998), the relationship between research funding and scientific production in nanotechnology (Beaudry and Allaoui, 2012), and the role of funding in nanotechnology impact (Wang and Shapira, 2011).

Zhao *et al.* (2018) suggested that research funding is an essential resource in the science reward system. Funded publications receive more citations than unfunded documents (Quinlan *et al.*, 2008; Yan *et al.*, 2018), and citation is positively associated with funding diversification and negatively associated with funding intensity (Gök *et al.*, 2015). However, some papers discussed the lack of funding and the consequences on research, especially on topics that are not perceived as having a high impact (Pendlebury, 1991).

Checchi *et al.* (2019) presented a performance-based research funding system and analyzed its effect on the quantity and quality of publications from different countries. The quality of research publications is associated with its funding status in medical education (Reed *et al.*, 2007).

Scholars argued that funding primarily affects citations, while funded articles attracted more usage, although with differences between scientific fields (Morillo, 2020; Pao, 1991). Ayoubi *et al.* (2019) stated that entering a grant competition, regardless of the outcome, helped scientists speed up their citations because of their efforts to prepare the proposal and communicate with other authors for the grant. Heyard *et al.* (2021) analyzed the relationship between funding and author metrics and showed that funded studies received more public attention than other research.

In general, the literature suggested that funded research has a more substantial impact than unfunded scientific research (Mosleh *et al.*, 2022) which is magnified by other factors such as collaboration or authorship affiliation (Rigby, 2013). Financial resources accelerated the diffusion of science and new knowledge to support the well-being of society (Laudel, 2006; Roshani *et al.*, 2021). However, some studies also suggested that there is no significant relationship between funding and publication performance in some research areas (Jacob and Lefgren, 2011).

Fleming *et al.* (2019) analyzed the impact of funding on innovation according to the performance of patents and concluded that federal support accelerates the production of innovation. And more recently, Doğramaci and Rossi-Fedele (2022) suggested that publication typology and journal impact factor were significant predictors of short-term social and academic impact of endodontic research articles. Non grant-funded research and the coverage in general news bulletins achieved higher social impact, while the social attention score was also strongly related to professional impact.

**Methodology**

The SDGs (United Nations, 2015) are targets for global development adopted by the United Nations in September 2015 and set to be achieved by 2030. They are 17 interconnected goals which constitute a universal call to action to end poverty, protect the planet and improve the lives and prospects of all the people on the planet.

Dimensions database provides a classification system for these 17 goals, most of which are interconnected. The automated classification of publications aligning to the goals is implemented by supervised machine learning based on extensive training sets and curated keyword searches. A detailed description of the article level classification system (rather than journal level classification system) developed by Digital Science and implemented in Dimensions can be found at Wastl *et al.* (2020).

The unit of study in this paper is the research article, and the source of data is Dimensions. Thus, the articles analyzed correspond to all those classified in this database into the SDGs. The source of altmetrics data in Dimensions is Altmetric. This is a current popular and one of the first altmetrics aggregator platforms, that originated in 2011 through the support of Digital Science. It tracks and accumulates mentions and views from different social media, news, blogs, and other platforms for scholarly articles. It also computes a weighted score called 'altmetric attention score' where each different category of mention contributes in a different way to the final score (Altmetric, 2021).

The altmetric attention score measures the social attention a paper is getting from mainstream and social media, public policy documents, Wikipedia, etc. It collects the online presence, and it allows to analyze the conversations around a particular paper. For clarity in exposing the results, in this paper we will refer to this measure as social attention score, or just like social attention.

In this paper we consider all research articles on the SDGs indexed in Dimensions in 2018, and citations and social attention scores counted in the period 2018-2021. Data was exported on April 29, 2022. A total of 367,704 research articles are analyzed, of which 123,451 are OA Gold (33.6%); 20,792 are OA Hybrid (5.7%); 31,639 are OA Green (8.6%); 31,866 are OA Bronze (8.7%); and 159,956 are Closed (43.5%). The total population of research articles in the database that same year was 4,081,634.

**Results and discussion**

The interest aroused by the 17 Sustainable Development Goals in the international scientific community can be measured by the production of publications in this regard. So, the prevalence of research on the SDGs in 2018, according to the Dimensions classification system, reached 9% of the total production of research articles collected in the Dimensions database. The main four goals in number of publications in 2018 are, in this order, 'Affordable and clean energy', 'Good health and well-being', 'Quality education', and 'Peace, justice and strong institutions' (Figure 1 and Table 1). These goals account for three quarters of the total articles production that year. Together with 'Climate action', these five goals bring 85% of the publications in 2018.

Regarding the financial support that these investigations receive, the prevalence of financed research varies in the interval 8.6 – 49.8% according to the SDGs (Figure 2 and Table 1). These minimum and maximum financing prevalence correspond to 'Quality education' and 'Affordable and clean energy', respectively. 'Peace, justice and strong institutions', 'Decent work and economic growth', and 'Gender equality' are also among the goals with the least funding, while 'Climate action', 'Life below water', and 'Life on land', are among the best funded. These data obtained for the SDGs confirm the inequalities in funding among fields that are observed in general research too. That is, the social sciences and humanities receive less funding than sciences in general, including the physical and life sciences.

Academic citation and social mention are influenced by many factors. Among these factors are the research field and funding. Citation and social attention do not correlate (Table 2) and, therefore, they measure different dimensions in the impact of research results. The Pearson linear correlation coefficients between times cited and social attention are all positive, although quite low for the most prevalent SDGs. There are important differences in correlation both in terms of funding and goal. The coefficients vary in the interval 0.09–0.79 depending on financing and SDGs. This means that both financing and field defined by the goal, affect the relationship between citation and social attention differently.

In general, the influence that research on the 17 SDGs has on subsequent scientific production does not coincide with the social attention that they arouse in society (Figure 3, Table 3 and Table 4 in the Appendix). Beyond the different measure units on which citation and social attention are counted, it is observed as the goals most cited by the researchers do not coincide with those that arouse greater social attention.

There are important differences both in terms of funding and SDGs. In general, citations vary in the interval 3.2–33 depending on financing and goal (Table 3). However, social attention does so in the interval 1.2–34.5. Therefore, financing and field affect the citation and social attention differently.

The average number of citations per publication varies according to fields between 3.2–17.3 in the case of unfunded research, while this average increases considerably in the case of funded research, varying in the interval 12.9–33. Moreover, the citation advantage due to funding varies between 91–343%, although in most cases (in 12 of 17 goals) the citation advantage due to funding ranges in 100–200%. The greatest citation advantages are obtained in some of the social sciences and humanities goals, in this order in 'Quality education', 'Peace, justice and strong institutions', and 'No poverty'. The least advantages are observed in 'Responsible consumption and production', and 'Climate action'.

On the other hand, the average social attention of a research varies according to fields between 1.2–12 in the case of unfunded research, while said average social attention increases considerably again in the case of funded research, reaching values in 3.6–34.5. Furthermore, the social attention advantage due to financing varies between 162–705%, although in most cases (in 14 of 17 goals) the social attention advantage due to financing is between 170–350%. The greatest advantages of social attention due to funding are also obtained in 'Quality education', 'No poverty', and 'Peace, justice and strong institutions.' The least advantages are observed in 'Affordable and clean energy', and 'Responsible consumption and production', with 'Climate action' also among those with the least social attention advantage.

This means that funded research on SDGs receives in most cases an average of two to three times more citations than unfunded research. These differences are even greater in social attention, with funded research on SDGs in most cases receiving an average of 2.5 to 4.5 times more social attention than unfunded research.

Considering the median as central tendency measure instead of the mean (Table 4 in the Appendix), the differences are even greater. In general, there are again important differences both in terms of funding and goal. The median citation per publication varies according to fields in the interval 0–5 in the case of unfunded research, while this median increases considerably in the case of funded research, varying between 7–20. Note that the ranges of variation for the median do not overlap each other. That is, regardless of the field, half of the publications without funding are cited less than 5 times. However, half of the funded publications are cited more than 7 times. Moreover, the citation advantage due to funding varies between 220–1100%, although in most cases (in 11 of 17 goals) the citation advantage due to funding ranges between 300–700%.

On the other hand, the median social attention of a research is zero for all fields in the case of unfunded research, while said median attention increases in the case of funded research, reaching values between 0–5. That is, more than half of the research without funding receives no social attention. However, except for four of the goals, more than half of the funded research does receive some social attention.

As indicated, many of the research articles in this study have not been cited or have not received any attention in social media (compiled in the database after four years). This is something that happens frequently in the literature in the case of the citations (Dorta-González *et al.*, 2020). This fact corresponds to many zeros in the citation and social attention distributions, which are therefore highly skewed toward zero. As see in Figure 4, the distributions are skewed, especially the social attention. Notice that the mean, represented by a cross, is much higher than the median, represented by the central line in the box. This means that a small proportion of articles accumulates a large part of the citations and, above all, most of the social attention.

A non-parametric Kruskal-Wallis test was performed to analyze the effect of funding on citations and social attention. We concluded that the differences between groups (funded and not funded) are significant at 0.01 for all sustainable development goals. Figure 4 shows how both citation and social attention grow with financing. Notice that funded research is on the left (in blue) and not funded is on the right (in red) of the chart. This trend is observed both in the mean and in the median, and even in the rest of the quartiles in the distribution represented by the boxes and whiskers in the chart.

Moreover, Figure 4 allows observe the differences in the orders of magnitude in the impacts among SDGs. For instance, research on 'Gender equality', 'Climate action' and 'Peace, justice and strong institutions' are generally those that receive the most social attention, with scores of the same order of magnitude as citation and means much higher than in the case of citation. However, there are goals that barely receive social attention, as is the case of goals 6, 7, 9, 11, and 12.

We next analyze the funding effect on citation and social attention according to the access type (see Table 5 in the Appendix). Regarding the form of access, there are significant differences in the average citation and social attention. These differences, meanwhile, are mostly due to the peculiarities of the access type, and relevant citation and social attention advantages are still frequently shown in favour of financed research.

In general, the modalities of access to publications with a greater citation advantage due to funding coincide with those that also have a greater advantage in social attention. And the modalities with the least advantage in citation due to funding are at the same time those with the least advantage in social attention. Furthermore, according to the range of variation of the mean in each modality, OA Gold and OA Hybrid are generally among the modalities with the greatest advantage in citation and social attention due to funding, while the access modalities with the least advantage in terms of financing are OA Green and Closed. The OA Bronze modality presents a variable behaviour in the data, but in general it is usually found in intermediate positions in terms of funding advantage both in citation and social attention. This may be because the decision on this modality corresponds to the publisher and not the authors.

Among unfunded research, in general, the most cited access modality and the one that receives most social attention is OA Green, while OA Gold is among the least cited and socially attended. In the case of the OA Green, starting from the highest reference values, the advantage of citation and social attention due to funding is therefore among the lowest, reaching negative in some cases. However, for the OA Gold, starting from the lowest reference values, the advantage of citation and social attention due to funding is therefore among the highest.

**Conclusions**

We quantified the contribution of funding to the academic citation and social attention of research. As a novel contribution, we analyzed the social attention and the effect of the access type to the publication, in particular the open access modalities. We considered the accumulated citation and social attention until the year 2021 of more than 367 thousand research articles published in the year 2018. The data source was Dimensions and the research articles were those about the Sustainable Development Goals.

The interest about the SDGs in the international scientific community in 2018 reached 9% of the total production of research articles according to the Dimensions classification system. Three quarters of the total articles production in that year were in this order about 'Affordable and clean energy', 'Good health and well-being', 'Quality education', and 'Peace, justice and strong institutions.'

The proportion of financed research varied between 9–50% according to the SDGs. These data obtained for the SDGs confirm the funding inequalities among fields that are observed in research. That is, the social sciences and humanities receive less funding than sciences.

Academic citation and social mention are influenced by many factors. Among these factors are the research field and funding. Citation does not correlate with social attention and, therefore, they measure different dimensions in the impact of research results. Moreover, the goals most cited by the researchers do not coincide with those that arouse greater social attention.

We observed that funded research on SDGs receives in most cases an average of two to three times more citations than unfunded research. These differences are even greater in social attention, with funded research on SDGs in most cases receiving an average of 2.5 to 4.5 times more social attention than unfunded research.

The average number of cites per article, four years after its publication, varies according to the goals between 3.2–17.3 in the case of unfunded research, while this average increases considerably in the case of funded research, reaching average citation in the interval 12.9–33. Moreover, the citation advantage due to funding varies in most cases between 100–200%. On the other hand, the average social attention of a research varies according to goals between 1.2–12 in the case of unfunded research, while it increases considerably again in the case of funded research, reaching values in 3.6–34.5. Furthermore, the social attention advantage due to funding varies in most cases between 170–350%. The greatest citation and social attention advantages due to funding are obtained in some of the social sciences and humanities goals. This is because only a small proportion of the highest-impact research is funded in these branches of knowledge.

Many research articles in this study did not receive any social attention or were not cited within four years of their publication. Furthermore, a small proportion of articles garnered most of the citations and, above all, most of the social attention.

Both citation and social attention grew with funding. Research on 'Gender equality', 'Climate action' and 'Peace, justice and strong institutions' received the greatest social attention, with scores of the same order of magnitude as citation and means much higher than in the case of citation. However, there were other goals that barely received social attention.

There were important differences in average citation and social attention regarding the type of access. However, these differences were mainly attributed to the characteristics of the access type, and relevant advantages of citation and social attention were observed toward funded research in all access types and most of the goals.

The journals with the greatest citation advantage because of funding generally had the same access options as those with the greatest social attention advantage. Similarly, those with a lower citation advantage also had a lower advantage in social attention. The access modes with the least advantage in terms of finance were OA Green and Closed, while OA Gold and OA Hybrid were often among those with the greatest advantage in citation and social attention.

There are several things to consider regarding the access typology. The OA Green was typically the access modality that received the most citations and social attention among the unfunded research, whereas the OA Gold was among the least cited and socially attended. The advantage of citation and social attention due to funding was thus one of the lowest, and in some cases even negative, in the case of OA Green because it starts from higher citation averages.

The citation and social attention advantage due to funding was, however, one of the biggest for OA Gold due to lower citation rates for unfunded research. This study used the altmetric attention score. This is a mixed indicator, aggregating multiple sources into a single score (Altmetric, 2021). Mixed indicators cannot be given a robust interpretation.

For this reason, it should be avoided when evaluating researchers, especially in recruitment processes and internal promotions. However, in this work this indicator has been used to assess the research communication process and not the researchers.

Altmetric indicators have the advantage to measure different types of impacts beyond academic citations. They also have the potential to capture earlier impact evidence. This is useful in self-assessments. This is also useful when studying science itself, as is the case in this study. Nevertheless, social attention should be used with caution because it could give a partial and biased view of all types of social impact, in addition to the fact that it does not differentiate positive from negative impact.

Regarding the practical implications, when funding is not available to publish in open access or hybrid journals, it is recommended to self-archive the pre-print or post-print version in an open access repository. And in relation to the social implications, albeit with caution, it is convenient to incorporate the social impact as one more dimension of all the impact generated by scientific research. In this way, it is possible to have a broader and more adjusted vision of the true impact of the research beyond the academic impact measured by the citations received.

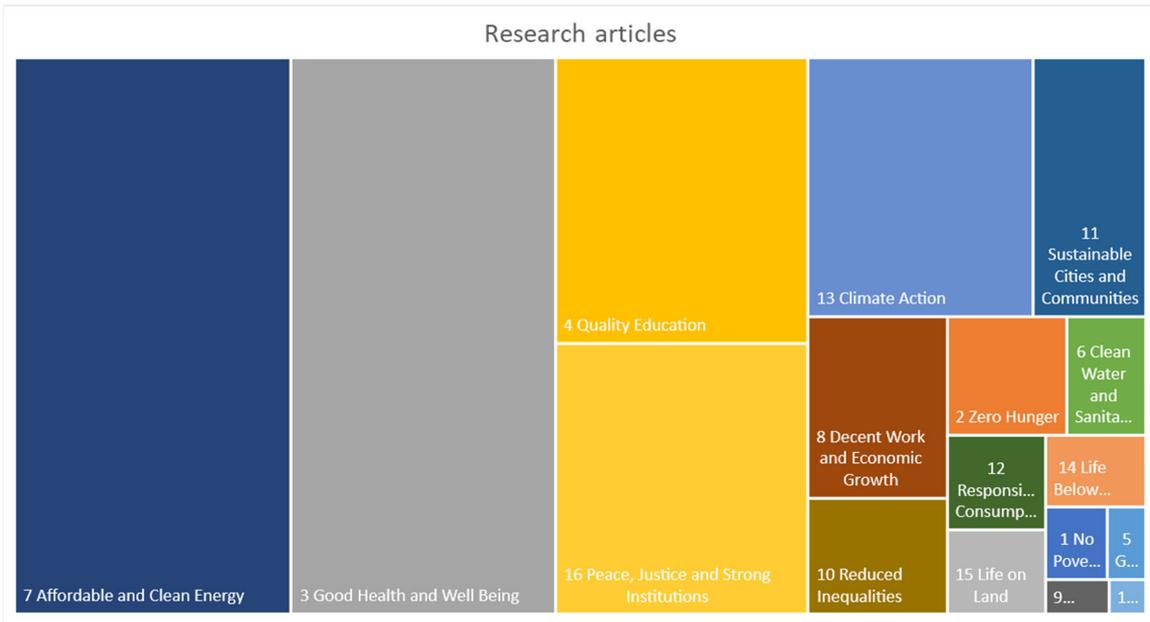

Figure 1. Treemap for prevalence of research in the SDGs. Research articles in 2018 and fields based on Dimensions classification system

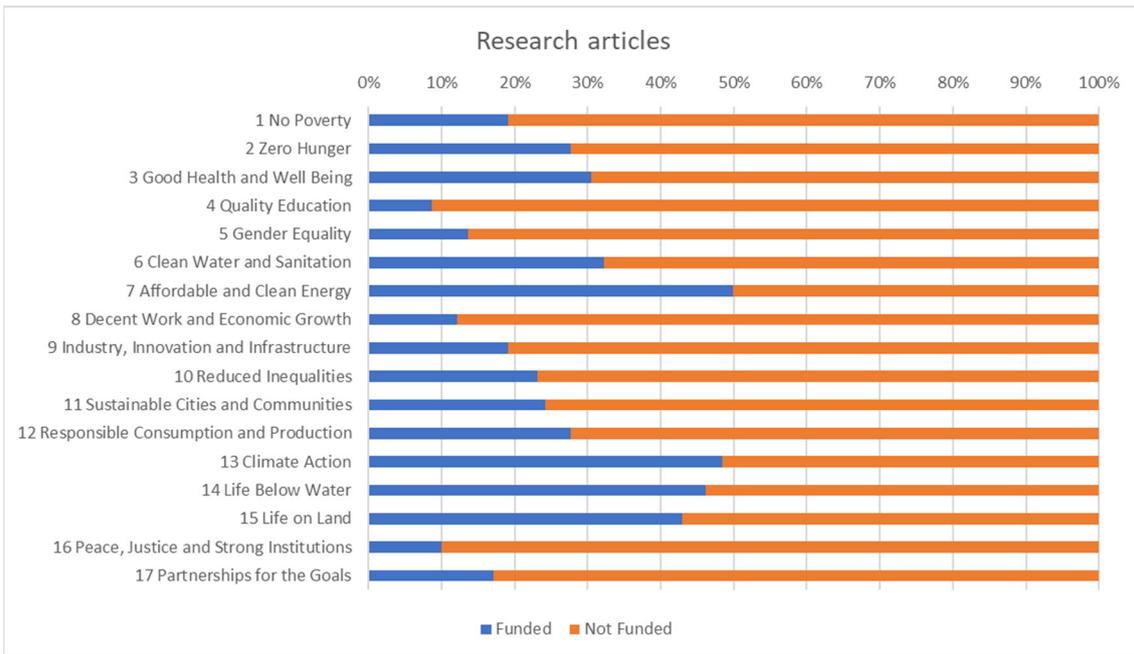

Figure 2. Prevalence of funded research in the SDGs. Research articles in 2018 and fields based on Dimensions classification system

Table 1. Prevalence of research in the SDGs by funding. Research articles in 2018 and fields based on Dimensions classification system

| UN Sustainable Development Goals | Funded | | Not Funded | | All | |
|---|---:|---:|---:|---:|---:|---:|
| | Articles | % | Articles | % | Articles | % of Total |
| 1 No Poverty | 529 | 19.1% | 2,245 | 80.9% | 2,774 | 0.8% |
| 2 Zero Hunger | 2,431 | 27.7% | 6,356 | 72.3% | 8,787 | 2.4% |
| 3 Good Health and Well-being | 27,761 | 30.4% | 63,463 | 69.6% | 91,224 | 24.8% |
| 4 Quality Education | 3,864 | 8.6% | 40,954 | 91.4% | 44,818 | 12.2% |
| 5 Gender Equality | 238 | 13.6% | 1,506 | 86.4% | 1,744 | 0.5% |
| 6 Clean Water and Sanitation | 1,881 | 32.2% | 3,955 | 67.8% | 5,836 | 1.6% |
| 7 Affordable and Clean Energy | 47,571 | 49.8% | 47,875 | 50.2% | 95,446 | 26.0% |
| 8 Decent Work and Economic Growth | 1,901 | 12.2% | 13,731 | 87.8% | 15,632 | 4.3% |
| 9 Industry, Innovation and Infrastructure | 262 | 19.2% | 1,106 | 80.8% | 1,368 | 0.4% |
| 10 Reduced Inequalities | 2,308 | 23.1% | 7,670 | 76.9% | 9,978 | 2.7% |
| 11 Sustainable Cities and Communities | 4,390 | 24.1% | 13,804 | 75.9% | 18,194 | 4.9% |
| 12 Responsible Consumption and Production | 1,595 | 27.6% | 4,176 | 72.4% | 5,771 | 1.6% |
| 13 Climate Action | 17,446 | 48.4% | 18,607 | 51.6% | 36,053 | 9.8% |
| 14 Life Below Water | 2,085 | 46.2% | 2,430 | 53.8% | 4,515 | 1.2% |
| 15 Life on Land | 2,206 | 43.0% | 2,926 | 57.0% | 5,132 | 1.4% |
| 16 Peace, Justice and Strong Institutions | 4,221 | 9.9% | 38,284 | 90.1% | 42,505 | 11.6% |
| 17 Partnerships for the Goals | 136 | 17.1% | 658 | 82.9% | 794 | 0.2% |
| *Total | | | | | 367,704 | 100% |
| Range of variation | | 8.6–49.8% | | 50.2–91.4% | | 0.2–26.0% |

\* Given the interdisciplinarity of the SDGs, some articles have been assigned by the classification system to more than one goal. Therefore, the sum does not match the total.

Table 2. Pearson linear correlation coefficient between times cited and social attention by funding and SDGs. Fields based on Dimensions classification system. Production in 2018 and impact in 2018-2021

| UN Sustainable Development Goals | Funded | Not Funded |
|---|---|---|
| 1 No Poverty | 0.66 | 0.27 |
| 2 Zero Hunger | 0.32 | 0.25 |
| 3 Good Health and Well-being | 0.18 | 0.24 |
| 4 Quality Education | 0.34 | 0.36 |
| 5 Gender Equality | 0.39 | 0.62 |
| 6 Clean Water and Sanitation | 0.38 | 0.30 |
| 7 Affordable and Clean Energy | 0.15 | 0.12 |
| 8 Decent Work and Economic Growth | 0.48 | 0.41 |
| 9 Industry, Innovation and Infrastructure | 0.09 | 0.13 |
| 10 Reduced Inequalities | 0.33 | 0.29 |
| 11 Sustainable Cities and Communities | 0.20 | 0.17 |
| 12 Responsible Consumption and Production | 0.10 | 0.13 |
| 13 Climate Action | 0.41 | 0.32 |
| 14 Life Below Water | 0.51 | 0.33 |
| 15 Life on Land | 0.28 | 0.45 |
| 16 Peace, Justice and Strong Institutions | 0.39 | 0.79 |
| 17 Partnerships for the Goals | 0.68 | 0.20 |
| **Range of variation** | 0.09–0.68 | 0.12–0.79 |

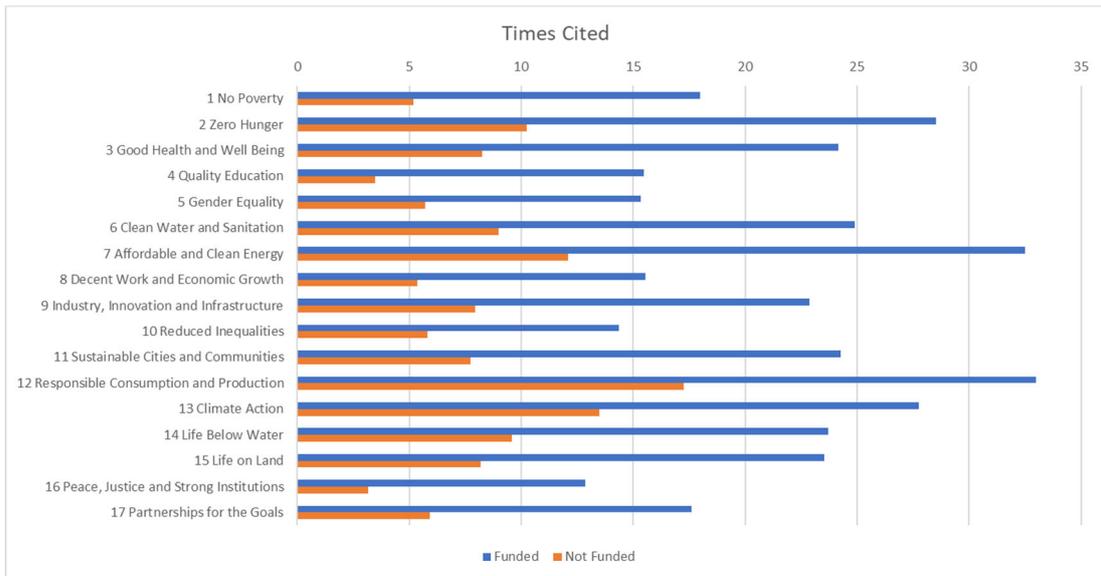
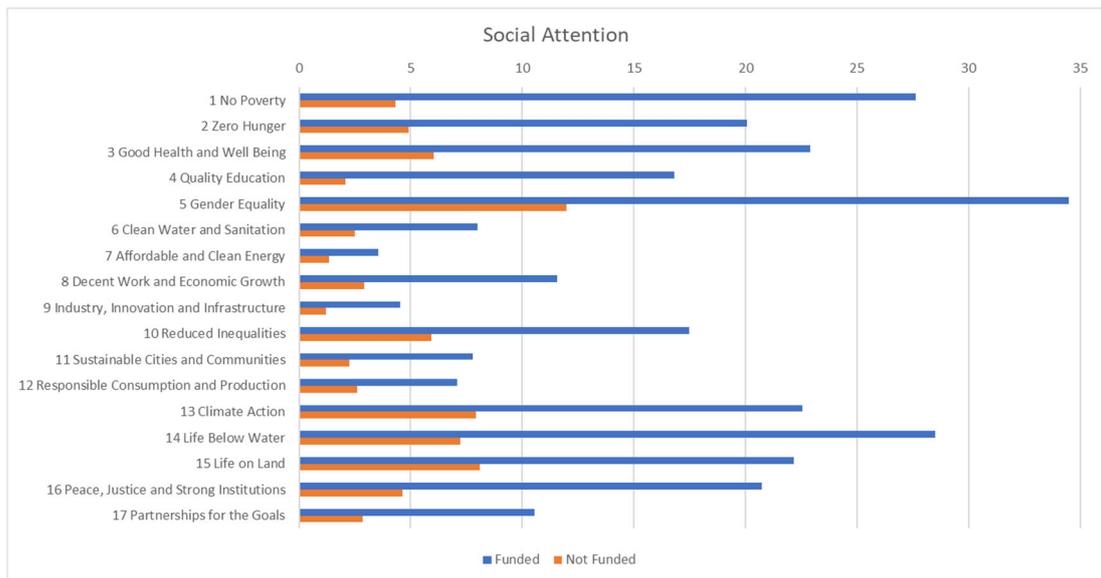
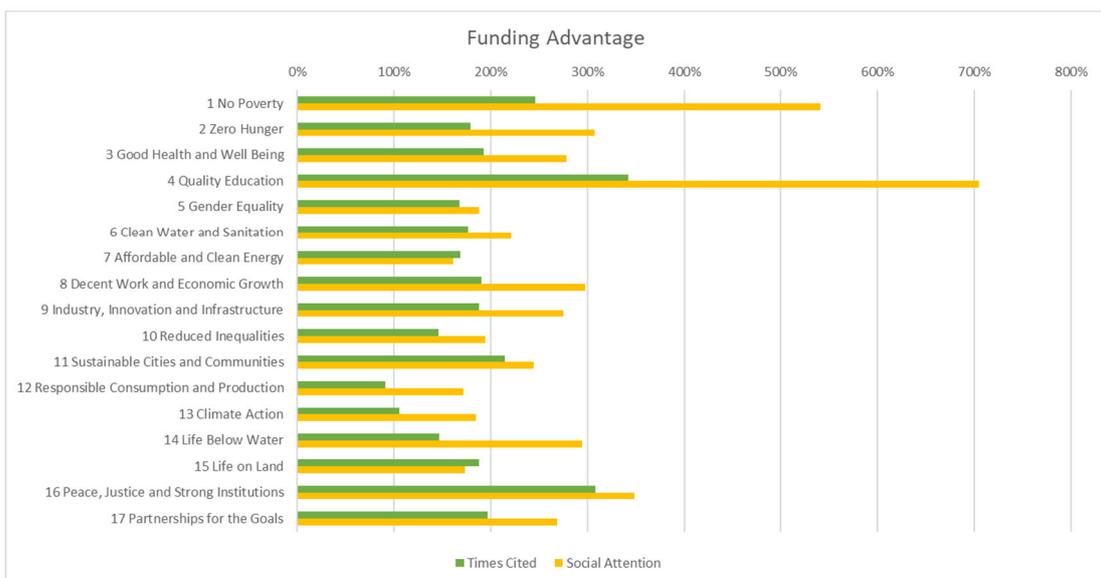

Figure 3. Mean of citation and social attention by funding and SDGs. Fields based on Dimensions classification system. Production in 2018 and impact in 2018-2021

Table 3. Mean of citation and social attention by funding and SDGs. Fields based on Dimensions classification system. Production in 2018 and impact in 2018-2021

| UN Sustainable Development Goals | Funded | | Not Funded | | Funding Advantage | |
|---|---|---|---|---|---|---|
| | Times cited | Social attention | Times cited | Social attention | Times cited | Social attention |
| 1 No Poverty | 18.0 | 27.6 | 5.2 | 4.3 | 246% | 541% |
| 2 Zero Hunger | 28.5 | 20.1 | 10.2 | 4.9 | 179% | 307% |
| 3 Good Health and Well-being | 24.2 | 22.9 | 8.3 | 6.0 | 193% | 279% |
| 4 Quality Education | 15.5 | 16.8 | 3.5 | 2.1 | 343% | 705% |
| 5 Gender Equality | 15.3 | 34.5 | 5.7 | 12.0 | 168% | 188% |
| 6 Clean Water and Sanitation | 24.9 | 8.0 | 9.0 | 2.5 | 177% | 221% |
| 7 Affordable and Clean Energy | 32.5 | 3.6 | 12.1 | 1.4 | 169% | 162% |
| 8 Decent Work and Economic Growth | 15.5 | 11.6 | 5.4 | 2.9 | 190% | 297% |
| 9 Industry, Innovation and Infrastructure | 22.9 | 4.5 | 7.9 | 1.2 | 188% | 275% |
| 10 Reduced Inequalities | 14.3 | 17.5 | 5.8 | 5.9 | 146% | 194% |
| 11 Sustainable Cities and Communities | 24.3 | 7.8 | 7.7 | 2.3 | 215% | 245% |
| 12 Responsible Consumption and Production | 33.0 | 7.1 | 17.3 | 2.6 | 91% | 172% |
| 13 Climate Action | 27.7 | 22.5 | 13.5 | 7.9 | 105% | 185% |
| 14 Life Below Water | 23.7 | 28.5 | 9.6 | 7.2 | 147% | 295% |
| 15 Life on Land | 23.6 | 22.2 | 8.2 | 8.1 | 188% | 174% |
| 16 Peace, Justice and Strong Institutions | 12.9 | 20.7 | 3.2 | 4.6 | 308% | 349% |
| 17 Partnerships for the Goals | 17.6 | 10.5 | 5.9 | 2.9 | 197% | 269% |
| **Range of variation** | 12.9–33.0 | 3.6–34.5 | 3.2–17.3 | 1.2–12.0 | 91–343% | 162–705% |

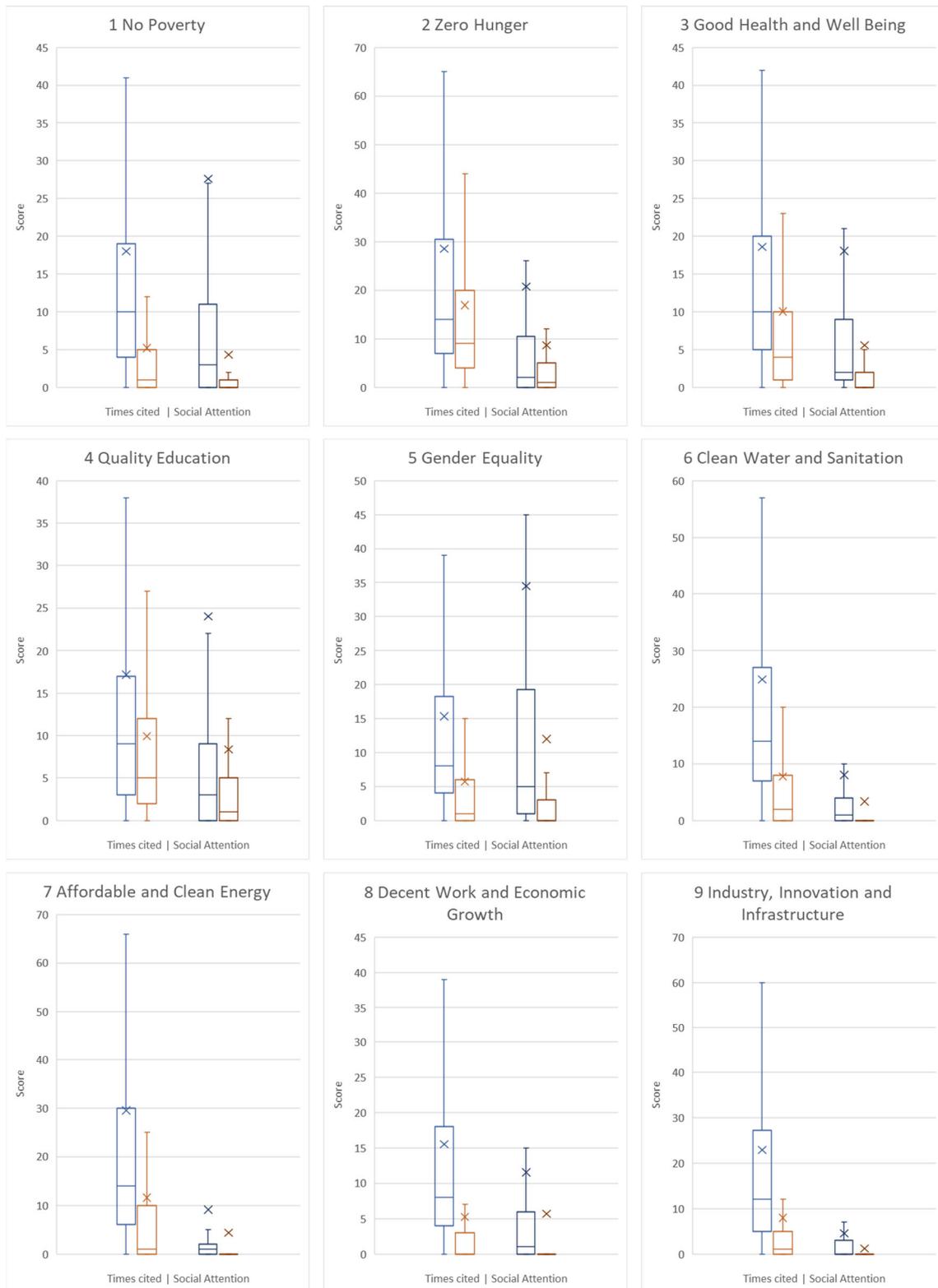

Figure 4. Score distribution for citation and social attention by funding and SDGs. Research funded on the left (in blue) and not funded on the right (in red) of the chart. Fields based on Dimensions classification system. Production in 2018 and impact in 2018-2021

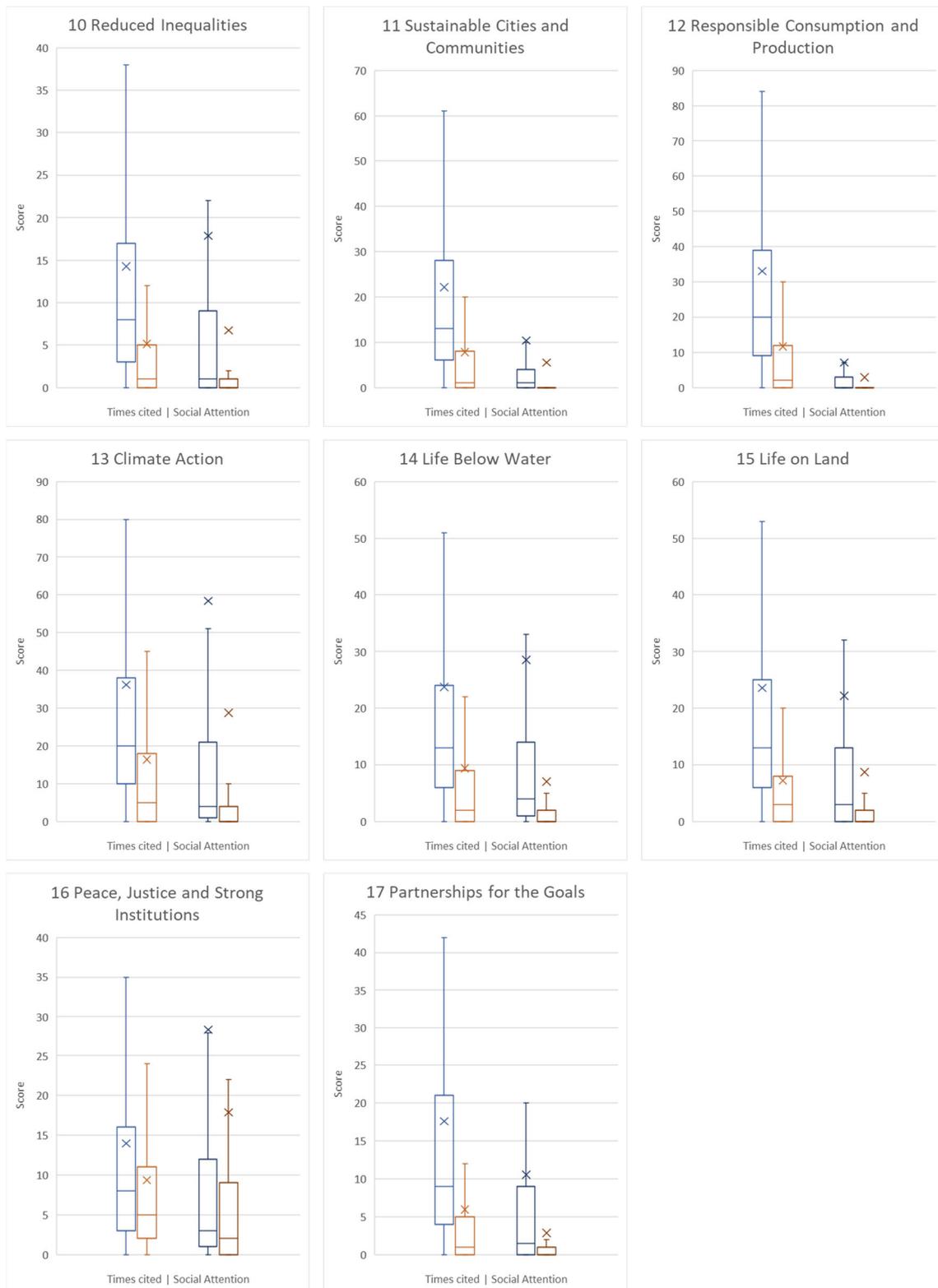

Figure 4. (Continuation) Score distribution for citation and social attention by funding and SDGs. Research funded on the left (in blue) and not funded on the right (in red) of the chart. Fields based on Dimensions classification system. Production in 2018 and impact in 2018-2021

**Appendix**

See Tables 4 and 5.

Table 4. Median of citation and social attention by funding and SDGs. Fields based on Dimensions classification system. Production in 2018 and impact in 2018-2021

| UN Sustainable Development Goals | Funded | | Not Funded | | Funding Advantage | |
|---|---|---|---|---|---|---|
| | Times cited | Social attention | Times cited | Social attention | Times cited | Social attention |
| 1 No Poverty | 10.0 | 3.0 | 1.0 | 0.0 | 900% | - |
| 2 Zero Hunger | 15.0 | 2.0 | 3.0 | 0.0 | 400% | - |
| 3 Good Health and Well-being | 12.0 | 2.0 | 2.0 | 0.0 | 500% | - |
| 4 Quality Education | 8.0 | 2.0 | 0.0 | 0.0 | - | - |
| 5 Gender Equality | 8.0 | 5.0 | 1.0 | 0.0 | 700% | - |
| 6 Clean Water and Sanitation | 14.0 | 1.0 | 2.0 | 0.0 | 600% | - |
| 7 Affordable and Clean Energy | 17.0 | 0.0 | 4.0 | 0.0 | 325% | - |
| 8 Decent Work and Economic Growth | 8.0 | 1.0 | 1.0 | 0.0 | 700% | - |
| 9 Industry, Innovation and Infrastructure | 12.0 | 0.0 | 1.0 | 0.0 | 1100% | - |
| 10 Reduced Inequalities | 8.0 | 1.0 | 2.0 | 0.0 | 300% | - |
| 11 Sustainable Cities and Communities | 15.0 | 0.0 | 2.0 | 0.0 | 650% | - |
| 12 Responsible Consumption and Production | 20.0 | 0.0 | 5.0 | 0.0 | 300% | - |
| 13 Climate Action | 16.0 | 1.0 | 5.0 | 0.0 | 220% | - |
| 14 Life Below Water | 13.0 | 4.0 | 3.0 | 0.0 | 333% | - |
| 15 Life on Land | 13.0 | 3.0 | 3.0 | 0.0 | 333% | - |
| 16 Peace, Justice and Strong Institutions | 7.0 | 3.0 | 0.0 | 0.0 | - | - |
| 17 Partnerships for the Goals | 9.0 | 1.5 | 1.0 | 0.0 | 800% | - |
| **Range of variation** | 7–20 | 0–5 | 0–5 | 0–0 | 220–1100% | – |

Table 5. Mean of citation and social attention by funding, access type, and SDGs. Fields based on Dimensions classification system. Production in 2018 and impact in 2018-2021

| UN Sustainable Development Goals | Access Type | Funded | | | Not Funded | | | Funding Advantage | |
|---|---|---|---|---|---|---|---|---|---|
| | | N | Times cited | Social attention | N | Times cited | Social attention | Times cited | Social attention |
| 1 No Poverty | OA Gold | 125 | 15.4 | 27.6 | 884 | 3.2 | 1.9 | 378% | 1381% |
| | OA Hybrid | 45 | 26.0 | 66.8 | 185 | 5.4 | 9.0 | 381% | 646% |
| | OA Green | 115 | 21.7 | 29.7 | 139 | 10.0 | 6.8 | 117% | 337% |
| | OA Bronze | 43 | 21.4 | 45.3 | 213 | 4.5 | 5.2 | 371% | 776% |
| | Closed | 201 | 15.0 | 13.9 | 824 | 6.6 | 5.3 | 125% | 165% |
| 2 Zero Hunger | OA Gold | 756 | 28.0 | 19.1 | 2648 | 9.2 | 4.8 | 204% | 297% |
| | OA Hybrid | 241 | 36.8 | 34.2 | 383 | 14.6 | 13.2 | 152% | 159% |
| | OA Green | 365 | 35.6 | 35.8 | 311 | 16.7 | 8.9 | 113% | 302% |
| | OA Bronze | 222 | 31.2 | 24.8 | 573 | 6.8 | 5.4 | 361% | 356% |
| | Closed | 847 | 23.0 | 8.9 | 2441 | 10.7 | 3.1 | 115% | 184% |
| 4 Quality Education | OA Gold | 926 | 12.8 | 14.8 | 19234 | 2.3 | 1.0 | 459% | 1390% |
| | OA Hybrid | 304 | 27.5 | 29.7 | 2386 | 3.0 | 1.7 | 825% | 1631% |
| | OA Green | 664 | 16.5 | 24.3 | 1759 | 9.3 | 9.0 | 78% | 170% |
| | OA Bronze | 371 | 26.7 | 31.0 | 3608 | 2.9 | 3.3 | 806% | 835% |
| | Closed | 1599 | 11.7 | 9.2 | 13967 | 4.7 | 2.5 | 151% | 270% |
| 5 Gender Equality | OA Gold | 49 | 13.8 | 40.0 | 494 | 3.6 | 5.2 | 288% | 671% |
| | OA Hybrid | 28 | 26.9 | 30.9 | 83 | 5.6 | 12.2 | 385% | 154% |
| | OA Green | 51 | 16.6 | 16.3 | 145 | 13.3 | 35.5 | 25% | -54% |
| | OA Bronze | 19 | 20.5 | 197.7 | 129 | 5.4 | 27.1 | 278% | 630% |
| | Closed | 91 | 10.8 | 8.7 | 655 | 5.8 | 8.9 | 88% | -2% |
| 6 Clean Water and Sanitation | OA Gold | 493 | 18.8 | 8.5 | 1635 | 5.7 | 1.9 | 232% | 346% |
| | OA Hybrid | 167 | 31.6 | 21.6 | 228 | 10.3 | 3.5 | 208% | 521% |
| | OA Green | 182 | 30.3 | 15.2 | 160 | 23.6 | 5.6 | 28% | 173% |
| | OA Bronze | 109 | 23.9 | 17.5 | 455 | 4.0 | 4.1 | 491% | 328% |
| | Closed | 930 | 26.0 | 2.8 | 1477 | 12.5 | 2.2 | 109% | 28% |
| 8 Decent Work and Economic Growth | OA Gold | 347 | 13.8 | 9.7 | 4959 | 3.0 | 1.2 | 360% | 677% |
| | OA Hybrid | 185 | 23.8 | 32.0 | 858 | 4.7 | 4.0 | 408% | 698% |
| | OA Green | 425 | 15.9 | 14.7 | 1240 | 11.0 | 5.5 | 44% | 167% |
| | OA Bronze | 119 | 18.8 | 26.6 | 1324 | 4.2 | 6.2 | 349% | 329% |
| | Closed | 825 | 13.8 | 4.0 | 5350 | 6.7 | 2.9 | 108% | 39% |
| 9 Industry, Innovation and Infrastructure | OA Gold | 56 | 20.7 | 6.0 | 483 | 4.7 | 0.7 | 342% | 750% |
| | OA Hybrid | 25 | 28.6 | 3.3 | 68 | 4.7 | 0.6 | 504% | 407% |
| | OA Green | 35 | 28.2 | 3.3 | 59 | 20.0 | 8.3 | 41% | -61% |
| | OA Bronze | 16 | 18.9 | 9.6 | 91 | 2.2 | 0.4 | 746% | 2537% |
| | Closed | 130 | 21.8 | 3.9 | 405 | 11.9 | 1.1 | 83% | 265% |
| 10 Reduced Inequalities | OA Gold | 530 | 13.3 | 14.3 | 2559 | 3.5 | 3.0 | 279% | 375% |
| | OA Hybrid | 243 | 19.1 | 38.1 | 411 | 7.9 | 8.6 | 143% | 342% |
| | OA Green | 536 | 14.7 | 17.3 | 856 | 9.1 | 9.6 | 61% | 81% |
| | OA Bronze | 209 | 15.3 | 51.2 | 634 | 6.5 | 9.3 | 134% | 452% |
| | Closed | 790 | 13.1 | 4.4 | 3210 | 6.4 | 6.3 | 105% | -30% |
| 11 Sustainable Cities and Communities | OA Gold | 1170 | 20.4 | 8.7 | 6427 | 4.8 | 1.3 | 322% | 569% |

|  | | | | | | | | |
|---|---|---|---|---|---|---|---|---|
|  | OA Hybrid | 330 | 31.9 | 17.3 | 719 | 7.2 | 2.5 | 345% | 590% |
|  | OA Green | 562 | 27.8 | 11.5 | 720 | 18.0 | 5.6 | 54% | 106% |
|  | OA Bronze | 221 | 18.6 | 17.1 | 935 | 2.5 | 3.1 | 633% | 453% |
|  | Closed | 2107 | 24.9 | 3.8 | 5003 | 11.0 | 2.8 | 126% | 35% |
| 12 Responsible Consumption and Production | OA Gold | 362 | 22.0 | 10.0 | 1625 | 8.9 | 1.7 | 148% | 475% |
|  | OA Hybrid | 114 | 47.2 | 25.5 | 161 | 14.9 | 8.2 | 218% | 211% |
|  | OA Green | 207 | 43.3 | 9.0 | 338 | 34.5 | 4.3 | 26% | 106% |
|  | OA Bronze | 76 | 31.4 | 12.6 | 253 | 4.6 | 6.1 | 578% | 105% |
|  | Closed | 836 | 33.4 | 2.3 | 1799 | 23.6 | 2.1 | 41% | 13% |
| 13 Climate Action | OA Gold | 3980 | 23.6 | 27.9 | 6611 | 8.1 | 4.8 | 191% | 479% |
|  | OA Hybrid | 1527 | 35.2 | 38.6 | 891 | 17.0 | 15.0 | 107% | 157% |
|  | OA Green | 2290 | 35.7 | 43.4 | 1293 | 23.7 | 20.4 | 51% | 113% |
|  | OA Bronze | 1498 | 26.3 | 30.4 | 1259 | 7.5 | 16.3 | 251% | 86% |
|  | Closed | 8151 | 26.4 | 9.6 | 8553 | 16.7 | 6.5 | 58% | 49% |
| 14 Life Below Water | OA Gold | 499 | 21.8 | 43.6 | 972 | 5.2 | 4.9 | 315% | 785% |
|  | OA Hybrid | 165 | 35.7 | 35.8 | 86 | 17.8 | 44.0 | 100% | -19% |
|  | OA Green | 282 | 22.3 | 34.4 | 149 | 26.8 | 15.7 | -17% | 120% |
|  | OA Bronze | 151 | 21.7 | 47.0 | 180 | 5.5 | 5.7 | 296% | 726% |
|  | Closed | 988 | 23.4 | 15.2 | 1043 | 11.2 | 5.4 | 108% | 182% |
| 15 Life on Land | OA Gold | 605 | 18.2 | 20.4 | 1306 | 6.1 | 4.4 | 196% | 366% |
|  | OA Hybrid | 180 | 36.1 | 47.5 | 188 | 10.6 | 13.6 | 242% | 248% |
|  | OA Green | 285 | 28.3 | 30.3 | 167 | 11.4 | 10.2 | 149% | 197% |
|  | OA Bronze | 180 | 47.9 | 40.4 | 244 | 7.0 | 20.8 | 588% | 94% |
|  | Closed | 956 | 18.6 | 12.7 | 1021 | 10.1 | 8.5 | 84% | 50% |
| 16 Peace, Justice and Strong Institutions | OA Gold | 745 | 13.0 | 24.9 | 13113 | 1.6 | 1.8 | 731% | 1271% |
|  | OA Hybrid | 424 | 18.0 | 33.4 | 2247 | 2.9 | 4.0 | 520% | 743% |
|  | OA Green | 1022 | 13.7 | 22.1 | 2496 | 7.4 | 11.2 | 85% | 97% |
|  | OA Bronze | 384 | 18.6 | 38.6 | 3609 | 3.3 | 9.4 | 472% | 309% |
|  | Closed | 1646 | 9.6 | 10.6 | 16819 | 3.8 | 4.9 | 155% | 116% |
| 17 Partnerships for the Goals | OA Gold | 42 | 18.1 | 11.2 | 271 | 3.8 | 2.2 | 380% | 404% |
|  | OA Hybrid | 17 | 13.5 | 14.8 | 44 | 3.4 | 3.2 | 298% | 364% |
|  | OA Green | 24 | 17.3 | 12.0 | 61 | 10.4 | 7.8 | 66% | 54% |
|  | OA Bronze | 13 | 14.5 | 6.0 | 49 | 4.0 | 2.4 | 262% | 147% |
|  | Closed | 40 | 20.0 | 8.6 | 233 | 8.2 | 2.3 | 145% | 270% |
| **Range of variation** | OA Gold |  | 12.8–28.0 | 6.0–43.6 |  | 1.6–9.2 | 0.7–5.2 | 148–731% | 297–1390% |
|  | OA Hybrid |  | 13.5–47.2 | 3.3–66.8 |  | 2.9–17.8 | 0.6–44.0 | 100–825% | -19–1631% |
|  | OA Green |  | 13.7–43.3 | 3.3–43.4 |  | 7.4–34.5 | 4.3–35.5 | -17–149% | -61–337% |
|  | OA Bronze |  | 14.5–47.9 | 6.0–197.7 |  | 2.2–7.5 | 0.4–27.1 | 134–806% | 86–2537% |
|  | Closed |  | 9.6–33.4 | 2.3–15.2 |  | 3.8–23.6 | 1.1–8.9 | 41–155% | -30–270% |

Note: The disaggregation of the data about objectives 3 and 7, by type of access, could not be done due to the considerable number of articles published in some typologies